\documentclass[a4paper,reqno,numberedheadings,12pt]{article}
\usepackage{graphicx,cite}
\usepackage{amsmath,ams thm,amssymb,dsfont,graphicx,rotating,setspace}
\newtheorem{theorem}{Theorem}[section]
\advance\textwidth 0.5in \advance\oddsidemargin -0.25in \advance\textheight 0.5in \advance\topmargin -0.5in

\def\bea{\begin{eqnarray}}
\def\eea{\end{eqnarray}}

   \def\({\left(} \def\){\right)}

\def\beq{\begin{equation}} \def\ee{\end{equation}} 

 \def\qed{\vrule height0.6em width0.3em depth0pt}

\title {\bf Lie point symmetries of differential--difference equations}

\author{{\bf D. Levi} \\ Dipartimento di Ingegneria Elettronica, \\ Universit\`a degli Studi Roma Tre and Sezione INFN, Roma Tre, \\ Via della
Vasca Navale 84, 00146 Roma, Italy \\ {\sl E-mail: levi@roma3.infn.it} \and {\bf P. Winternitz} \\
 Centre de recherches math\'{e}matiques et \\ D\'epartement de math\'ematiques et
statistique, Universit\'{e} de Montr\'{e}al \\ C.P.
6128,
succ. Centre--ville, H3C 3J7, Montr\'{e}al (Qu\'{e}bec), Canada\\ {\sl E-mail: wintern@crm.umontreal.ca} \and {\bf R.I. Yamilov} \\ Ufa Institute of Mathematics, Russian Academy
of Sciences, \\ 112 Chernyshevsky Street, Ufa 450008, Russian Federation \\ {\sl E-mail: RvlYamilov@matem.anrb.ru}}

\date{\today}
\begin{document}
\maketitle

\begin{abstract}
We present an algorithm for determining the Lie point symmetries of
differential equations
on fixed non transforming lattices, i.e. equations involving both
continuous and discrete independent variables. The symmetries of a
specific integrable discretization of the Krichever-Novikov equation, the
Toda lattice and Toda field theory are presented as examples of the
general method.

 \end{abstract}


\section{Introduction}\label{s1}
Two different but equivalent infinitesimal formalisms exist for calculating Lie point symmetries of differential equations \cite{r1}. One is that of `standard` vector fields
\bea \label{1.1}
\hat X = \sum_{i=1}^p \xi_i(\vec x, \vec u)\partial_{x_i} + \sum_{\alpha=1}^q \phi_{\alpha}(\vec x, \vec u)\partial_{u_{\alpha}}
\eea
acting on the independent variables $x_i$ and dependent ones $u_{\alpha}$ in the considered differential equation.

The other is that of the {\it evolutionary} vector fields
\bea \label{1.2}
\hat X^E = \sum_{\alpha=1}^q \mathcal Q_{\alpha}(\vec x, \vec u, {\vec u}_{\vec x})\partial_{u_{\alpha}},
\eea
acting only on the dependent variables.

The equivalence of the two formalisms is due to the fact that the total derivatives $D_{x_i}$ are themselves `generalized` symmetry operators, so for any differential equation
\bea \label{1.3}
\mathcal E(x_i, u_{\alpha}, u_{\alpha,x_i}, \cdots) = 0
\eea
we have
\bea \label{1.4}
\mbox{pr} \hat X^E \mathcal E |_{\mathcal E = 0} = (\mbox{pr} \hat X - \sum_{i=1}^p \xi_i D_{x_i}) \mathcal E |_{\mathcal E = 0}=0.
\eea
Here $\mbox{pr}\hat X^E$ and $\mbox{pr} \hat X$ are the appropriate differential prolongations of $\hat X^E$ and $\hat X$. Relation (\ref{1.4}) implies that for point transformations we have
\bea \label{1.5}
Q_{\alpha} = \phi_{\alpha} - \sum_{i=1}^p \xi_i u_{\alpha,x_i}.
\eea
For all details we refer to e.g. P. Olver`s textbook \cite{r1}.

An advantage of the standard formalism is its direct relation to the group transformations obtained by integrating the equations
\bea \label{1.6}
\frac{d \tilde x_i}{d \lambda}&=&\xi_i(\tilde {\vec x}, \tilde {\vec u}), \qquad \frac{d \tilde u_{\alpha}}{d \lambda}=\phi_{\alpha}(\tilde {\vec x}, \tilde {\vec u}), \\ \nonumber
\tilde { x_i}|_{\lambda=0}&=& x_i,\qquad  \tilde { u_{\alpha}}|_{\lambda=0}= u_{\alpha}, \quad i=1,\cdots, p, \, \alpha=1, \cdots, q.
\eea

One advantage of the evolutionary formalism is its direct relation to the existence of flows commuting with the studied equation (\ref{1.3})
\bea \label{1.7}
\frac{d \tilde u_{\alpha}}{d \lambda}=Q_{\alpha},
\eea
where $Q_{\alpha}$ is the characteristic of the vector field as in (\ref{1.5}).

Another advantage is that the evolutionary formalism can easily be adapted to the case of higher symmetries.

Let us now consider a purely discrete equation, i.e. a difference equation. We restrict to the case of one scalar function defined on a two dimensional lattice $u_{mn}$. We shall view $u$ as a continuous variable, introduce two further continuous variables $x$ and $t$ and consider $(x,t,u)$ as being evaluated, or sampled at discrete points on a lattice labelled by the indices $m$, $n$. We shall write $(x_{mn}, t_{mn}, u_{mn})$ for values at these points.

A difference system will consist of relations
\bea \label{1.8}
\mathcal E_a(x_{m+k,n+l}, t_{m+k,n+l}, u_{m+k,n+l})=0, \, a=1, \cdots, A \quad k_m \le k \le k_M, \, l_m \le l \le l_M,
\eea
between the variables $x$, $t$, and $u$ evaluated at a finite number of points on a lattice.

A Lie point symmetry of the system (\ref{1.8}) will be generated by a vector field
of the form
\bea \label{1.9}
\hat X_{mn} = \xi(x_{mn}, t_{mn}, u_{mn}) \partial_{x_{mn}} + \tau(x_{mn}, t_{mn}, u_{mn}) \partial_{t_{mn}}+\phi(x_{mn}, t_{mn}, u_{mn}) \partial_{u_{mn}}.
\eea

We see that the vector field (\ref{1.9}) for difference equations has the same form as (\ref{1.1}) for differential ones. Its prolongation is however different, namely
\bea \label{1.10}
\mbox{pr}\hat X = \sum_{k,l} \hat X_{m+k,n+l},
\eea
where the sum is over all points figuring in the system (\ref{1.8}).

In the continuous limit the system (\ref{1.8}) goes into a partial differential equation, eq. (\ref{1.10}) goes into the usual prolongation of a standard vector field (i.e. it also acts on functions of derivatives).

For recent reviews of the theory of continuous symmetries of difference equations see Ref. \cite{r2,r3,r31}.

The purpose of this article is to consider an intermediate case, that of differential--difference equations. In Section \ref{s2} we shall take a `semicontinuous` limit, i.e. leave the variable $x$ discrete but let $t$ tend to a continuous variable. This will provide us with both a standard and an evolutionary formalism for calculating point symmetries of differential--difference equations. In Section \ref{s3} we consider several special cases and prove some theorems that greatly simplify the calculation of symmetries. Section \ref{s4} is devoted to examples and Section 5 to a summary of the results obtained.

\section{Lie point symmetries of difference systems and their semicontinuous limit}\label{s2}

\subsection{The semicontinuous limit}
\begin{figure*}
\begin{center}
\includegraphics[scale=1]{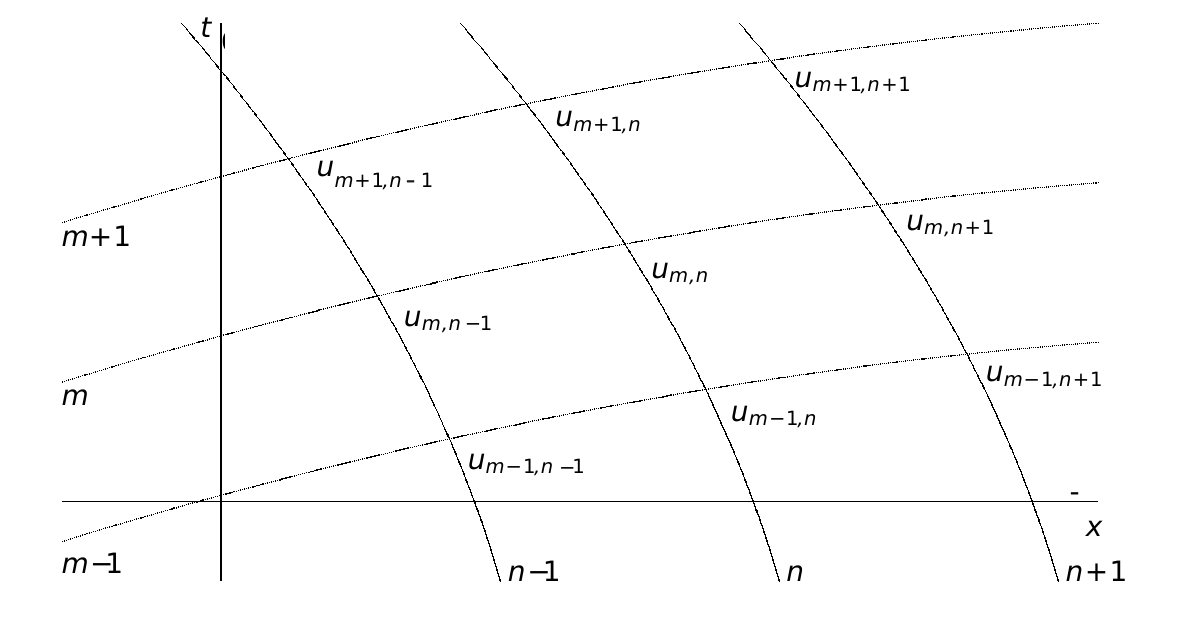}
\caption{\small Example of a two-dimensional lattice.}
\label{fig:endelmol}
\end{center}
\end{figure*}

A difference system is defined on a {\it discrete jet space}, a space of independent and dependent variables on a lattice.  In this article we restrict to the case of two independent variables $x$, $t$ and one dependent one $u$ defined on a two dimensional lattice with points labelled by two indices. We shall write
\bea \label{2.1}
\{ x_{mn}, t_{mn}, u_{mn} \equiv u(x_{mn}, t_{mn}) \},
\eea
(see Fig. \ref{fig:endelmol}).
\bigskip

\bigskip

The {\it discrete jet space} will be the set of all variables $\{ x_{jk}, t_{jk}, u_{jk} \}$ on the lattice. The dependence of $x$, $t$ and $u$ on the labels $m$, $n$ is determined from the difference system
\bea \label{2.2}
\mathcal E_{mn}^a (x_{jk}, t_{jk}, u_{jk}) = 0, \quad a=1,\cdots,N,
\eea
and some boundary conditions. In eq. (\ref{2.2}) $N$ is an integer satisfying $N \ge 3$ and $(j,k)$ run over some finite set of values on the lattice while $(m,n)$ is a fixed reference point. Eq. (\ref{2.2}) thus determines both the difference equation and the lattice.

Lie point symmetries of the system (\ref{2.2}) are generated by vector fields of the form
\bea \label{2.3}
\hat X_{mn}^D = \xi(x_{mn}, t_{mn}, u_{mn}) \partial_{x_{mn}} + \tau (x_{mn}, t_{mn}, u_{mn}) \partial_{t_{mn}}+\phi (x_{mn}, t_{mn}, u_{mn}) \partial_{u_{mn}}
\eea
(the superscript $D$ stands for `discrete`) satisfying
\bea \label{2.4}
\mbox{pr} \hat X^D \mathcal E^a |_{\mathcal E^1 =\mathcal E^2 = \cdots = \mathcal E^N = 0} = 0.
\eea
In eq. (\ref{2.4}) $\mbox{pr} \hat X^D$ is the prolongation of the vector field $\hat X_{mn}^D$ to the discrete jet space
\bea \label{2.5}
\mbox{pr} \hat X^D = \sum_{j,k} \hat X_{jk}^D,
\eea
where the sum is over all points figuring in the difference system (\ref{2.2}).

In this approach the lattice $(x_{mn}, t_{mn})$ is in general determined together with $u_{mn}$ from the system (\ref{2.2}) and the group transformations generated by the vector field $\hat X_{mn}^D$ also transform the lattice.  A special case corresponds to an a priori determined nontransforming lattice. In that case two of the equations in the system (\ref{2.2}) have the form
\bea \label{2.6}
x_{mn} = f(m,n), \qquad t_{mn} = g(m,n),
\eea
where $f$ and $g$ are given. Such is the case of a uniform orthogonal lattice where (\ref{2.6}) takes the form
\bea \label{2.7}
x_{mn} = \sigma_1 n + x_0, \qquad t_{mn} = \sigma_2 m + t_0,
\eea
and the scale factors $(\sigma_1, \sigma_2)$ and the origin $(x_0, t_0)$ are given numbers (e.g. $\sigma_1=\sigma_2=1, \, x_0=t_0=0)$.

We are interested in obtaining the form of the vector field in the semicontinuous limit in which $t_{mn}$ becomes a continuous variable $t$, but $x_{mn}$ remains discrete ( and independent of $t$). Thus $x$ will depend on one discrete label $n$ only and in particular $x$ may be given as $x_n = f(n)$, with $f(n)$ known (e.g. $x_n=hn$ for a uniform lattice, or $x_n = \lambda^n$ for an exponential one).

In this limit the difference system (\ref{2.2}) will reduce to a differential--difference equation (D$\Delta$E)
\bea \label{2.8}
\mathcal E_n(t, u_j, \dot u_j, \ddot u_j, \cdots, u_j^{(K)}) = 0, \quad n-L \le j \le n+M,
\eea
where dots denote $t$--derivatives, and $K$, $L$ and $M$ are some nonnegative integers. Together with eq. (\ref{2.8}) we have another equation which determines $x_n=f(n)$. We shall consider the case when $f(n)$ is already given (a known function) so that we can replace the dependence on $x_n$ by a dependence on the integer $n$ (without necessarily assuming that $f(n)$ is linear).

Eq. (\ref{2.8}) is thus defined on a `semidiscrete jet space` with local coordinates
\bea \label{2.9}
\{ t, u_j, \dot u_j, \ddot u_j, \cdots , u_j^{(K)}\},
\eea
where $j$ runs over all values on a one--dimensional lattice.

The vector field generating symmetries of eq. (\ref{2.8}) will have the form
\bea \label{2.10}
\hat X^{SD} = \tau_n ( t, u_n ) \partial_t + \phi_n( t, u_n ) \partial_{u_n},
\eea
($SD=$ semidiscrete) and its prolongation will be defined on the semidiscrete jet space (\ref{2.9}).

Let us consider the simplest nontrivial case, namely that of a difference system (\ref{2.2}) involving the three points $(m,n)$, $(m+1,n)$ and $(m,n+1)$, i.e. relating the variables $x_{jk}$, $t_{jk}$ and $u_{jk}$ in these 3 points:
\bea \label{2.11}
\mathcal E_n ( x_{mn}, x_{m+1,n}, x_{m,n+1},  t_{mn}, t_{m+1,n}, t_{m,n+1},  u_{mn}, u_{m+1,n}, u_{m,n+1}) = 0.
\eea

Before taking the limit we change notation and transform to new variables. We choose a reference point $(m,n)$ on the lattice (see Fig.1) and measure $t$ and $x$ from this point:
\bea \label{2.12}
t_{m+a,n+b} = t + \epsilon_{ab}, \quad x_{m+a,n+b} = f(n+b) + \theta(\epsilon_{ab}),
\eea
where $f(n+b)$ is a given function and $\theta(\epsilon_{ab})$ is an analytic function. Instead of $u_{m+j,n+k}$ we introduce a function $v_n(t)$
\bea \label{2.13}
v_{n+b}(t+\epsilon_{ab}) \equiv u_{m+a,n+b}(x_{m+a,n+b}, t_{m+a,n+b}),
\eea
and assume that the dependence on $t$ is analytical. Thus the dependence on $x$ (which remains discrete in the limit $\epsilon_{ab} \rightarrow 0$) is replaced by a dependence on the label $n+b$. For the reference point $x_{mn}$, $t_{mn}$ we put $\epsilon_{00}=0$, $\theta(\epsilon_{00})=0$.

So, in the case of the stencil $(m,n)$, $(m+1,n)$ and $(m,n+1)$ we have
\bea \nonumber
&&t_{mn}\equiv t,\qquad  t_{m+1,n} \equiv t +\epsilon_{10}, \qquad t_{m,n+1} \equiv t + \epsilon_{01},
\\ \nonumber
&&x_{mn}\equiv f(n), \qquad x_{m+1,n}=f(n) + \theta(\epsilon_{10}), \qquad x_{m,n+1} \equiv f(n+1)+ \theta(\epsilon_{01}),
\\ \label{2.14}
&& u_{mn}(x_{mn}, t_{mn}) = v_n(t), \qquad u_{m+1,n} \equiv v_n(t) + \epsilon_{10} v_{n,t}(t),
\\ \nonumber
&&u_{m,n+1}(x_{m,n+1}, t_{m,n+1}) = v_{n+1}(t + \epsilon_{01}) = v_{n+1}(t) + \sum_{j=1}^{\infty} \frac{\epsilon_{01}^j}{j!} v_{n+1}^{(j)}(t),
\eea
where $v_{n+1}^{(j)}(t)=\frac{d^j v_{n+1}(t)}{dt^j}$ and $v_{n,t}$ is the `discrete derivative` of $v_n(t)$ given by $v_n(t)= \frac{u_{m+1,n}-u_{m,n}}{\epsilon_{10}}$.

Since $v_n(t)$ is by assumption analytical, the Taylor series in (\ref{2.14}) are convergent. Using eq. (\ref{2.14}) we can also express the derivatives $\{ \partial_{t_{mn}}$, $\partial_{t_{m+1,n}}$, $\partial_{t_{m,n+1}}$, $\partial_{u_{mn}}$, $\partial_{u_{m+1,n}}$, $\partial_{u_{m,n+1}}\}$ in terms of $\{ \partial_t$, $\partial_{\epsilon_{10}}$, $\partial_{\epsilon_{01}}$, $\partial_{v_n}$, $\partial_{v_{n,t}}$, $\partial_{v_{n+1}}\}$ and thus transform the prolongation of the vector field (\ref{2.5}). We obtain
\bea \label{2.15}
\mbox{pr} \hat X^D &=& \tau_{mn} \partial_t + \phi_{mn} \partial_{v_n}+ (\tau_{m+1,n} - \tau_{mn}) \partial_{\epsilon_{10}} +  (\tau_{m,n+1} - \tau_{mn}) \partial_{\epsilon_{01}}
\\ \nonumber
&+& \Bigl [(\phi_{m+1,n} - \phi_{mn}) + (\tau_{mn}-\tau_{m+1,n}) v_{n,t} \Bigr ] \frac{1}{\epsilon_{10}}\partial_{v_{n,t}}
\\ \nonumber
&+& \Bigl [ \phi_{m,n+1} + (\tau_{mn}-\tau_{m,n+1}) \sum_{j=1}^{\infty} \frac{(t_{m,n+1}-t_{mn})^{j-1}}{(j-1)!} v_{n+1}^{(j)} \Bigr ] \partial_{v_{n+1}} .
\eea
Further, we put
\bea \label{2.16}
\tau_{mn} \equiv \tau_n(t,v_n), \qquad \phi_{mn} \equiv \phi_n(t,v_n),
\eea
and expand $\tau_{m+1,n}$ and $\phi_{m+1,n}$ about $\epsilon_{10}=0$, $\tau_{m,n+1}$ and $\tau_{m,n+1}$ and $\phi_{m,n+1}$ about $\epsilon_{01}=0$ and then let  $\mbox{pr}\hat X^D$  act on functions
\bea \label{2.20}
\mathcal E_n = \mathcal E_n(t, v_n, v_{n+1},  v_{n,t}),
\eea
obtained as  the limit of eq. (\ref{2.11}).
 In the semicontinuous limit we take $\epsilon_{10} \rightarrow 0$, $\epsilon_{01} \rightarrow 0$ and we obtain
\bea \label{2.17}
\lim_{(\epsilon_{10},\epsilon_{01})\rightarrow (0,0)} \mbox{pr} \hat X^D &=& \mbox{pr} \hat X^{SD} = \tau_n \partial_t + \phi_n \partial_{v_n} + \phi_n^{[t]}\partial_{ v_{n,t}} + \phi_n^{[n+1]}\partial_{ v_{n+1}^{(1)}},
\\ \label{2.18}
\phi_n^{[t]} &=& D_t \phi_n - (D_t \tau_n) \dot v_n,
\\ \label{2.19}
\phi_n^{[n+1]} &=& \phi_{n+1} + (\tau_n - \tau_{n+1}) v_{n+1}^{(1)}.
\eea

The form (\ref{2.18}) of the coefficient $\phi_n^{[t]}$ is the "obvious" generalization of the first prolongation for ordinary differential equations. The presence of the second term in (\ref{2.19}) is less obvious and follows from the above analysis of the semicontinuous limit. We see that the prolongation of the vector field $\hat X^{SD}$ to derivatives is the same as for differential equations  The prolongation to other points $x_n$ on the lattice does however {\bf not} consist of merely shifting $n$ in $\phi_n$.

We mention that the additional term in $\phi_n^{[n+1]}$ was missed in the article  \cite{r4}.

If we start from the set of all 9 points on the stencil of Fig.1 and take the semicontinuous limit in the same way, we arrive at a more general D$\Delta$E, namely
\bea \label{2.21}
\mathcal E_n(t, u_n, u_{n+1}, u_{n-1}, \dot u_n, \dot u_{n+1}, \dot u_{n-1}, \ddot u_n, \ddot u_{n+1}, \ddot u_{n-1} ) = 0,
\eea
(with the change of notation to $u_n$). We also obtain the prolongation $\mbox{pr} \hat X^{SD}$ for such an equation (see below).

\subsection{The Evolutionary Formalism and Commuting Flows for Differential--Difference Equations}

An alternative method of calculating symmetries of D$\Delta$E on a fixed lattice is to construct commuting flows in two variables. Let us again consider  eq. (\ref{2.20}), this time solved for the first derivative, and change the notation from $\dot v_n$ to $u_{n,t}$, which now denotes an ordinary time derivative;
\bea \label{2.22}
\dot u_n \equiv u_{n,t} = \mathcal F_n (t, u_n, u_{n+1}).
\eea

We introduce an additional variable $\lambda$, the group parameter and consider the flow on $u_n(t,\lambda)$ in this variable
\bea \label{2.23}
u_{n,\lambda} = \mathcal Q_n(t,u_n, \dot u_n).
\eea

Let us now require that the flows (\ref{2.22}) and (\ref{2.23}) be compatible, i.e. commute.  Thus we impose
\bea \label{2.24}
u_{n,t \lambda} = u_{n, \lambda t}.
\eea
We replace $u_{n,\lambda}$ using eq. (\ref{2.23}), $\dot u_{n}$ and $\ddot u_{n}$ using (\ref{2.22}) and its differential consequences and obtain
\bea \label{2.25}
\mathcal Q_{n,t}&+& \mathcal Q_{n,u_n} \mathcal F_n + \mathcal Q_{n,\dot u_n} \bigl( \mathcal F_{n,t} + \mathcal F_{n,u_n} \mathcal F_n + \mathcal F_{n,u_{n+1}} \mathcal F_{n+1} \bigr ) =
\\ \nonumber
&&\mathcal F_{n,u_n} \mathcal Q_n+\mathcal F_{n,u_{n+1}} \mathcal Q_{n+1}.
\eea

This derivation of (\ref{2.25})is completely equivalent to the following procedure. We first introduce an evolutionary vector field
\bea \label{2.26}
\hat X_E = \mathcal Q_n(t, u_n, \dot u_n, \cdots) \partial_{u_n},
\eea
and its prolongation
\bea \label{2.27}
\mbox{pr} \hat X_E = \mathcal Q_n \partial_{u_n} + \mathcal Q_{n+1} \partial_{u_{n+1}} + (D_t \mathcal Q_n)\partial_{\dot u_n} + \cdots .
\eea
We then apply this prolonged field to eq. (\ref{2.22}), require
\bea \label{2.28}
\mbox{pr} \hat X_E \bigl[ \dot u_n - \mathcal F_n(t, u_n, u_{n+1}) \bigr] \Bigl |_{(\dot u_n = \mathcal F_n, \ddot u_n = D_t \mathcal F_n)} = 0,
\eea
and reobtain eq. (\ref{2.25}).

Let us now specialize to the case of point symmetries. The quantity $\mathcal Q_n$ in (\ref{2.23}) and (\ref{2.26}) is the {\it characteristic} of the vector field $\hat X_E$. For point symmetries it has the form
\bea \label{2.29}
\mathcal Q_n(t, u_n, \dot u_n, \cdots) = \phi_n(t,u_n) - \tau_n(t, u_n) \dot u_n.
\eea
The total derivative $D_t$ is itself a (generalized) symmetry of the D$\Delta$E (\ref{2.8}) and in particular (\ref{2.22}). This provides us with a relation between ordinary and evolutionary vector fields and their prolongations, namely
\bea \label{2.30}
\mbox{pr} \hat X = \mbox{pr} \hat X_E + \tau_n(t,u_n) D_t.
\eea
Putting (\ref{2.29}) and (\ref{2.27}) into (\ref{2.30}) we reobtain eqs. (\ref{2.17}--\ref{2.19}).

We see that the "obvious" prolongation (\ref{2.27})of the evolutionary vector field (\ref{2.26}) provides, via eq. (\ref{2.30}) the correct prolongation (\ref{2.17}) of the ordinary vector field (\ref{2.10}).

\subsection{General Algorithm for Calculating Lie Point Symmetries of a Differential Difference Equation}

Let us consider a D$\Delta$E involving $L+M+1$ points and $t$ derivatives up to order $K$ as in eq. (\ref{2.8}). The Lie point symmetries of eq. (\ref{2.8}) can be obtained using the evolutionary formalism by imposing
\bea \label{2.31}
\mbox{pr}\hat X_E \mathcal E_n \bigl |_{\mathcal E_n=0,\, D^k_t \mathcal E_n=0} = 0, \qquad k=1, \cdots, K.
\eea
Thus the expression $\mbox{pr}\hat X_E \mathcal E_n$ is anihilated on the solution set of the equation (\ref{2.8}) and of all differential consequences of the equation.

The vector field $\hat X_E$ has the form (\ref{2.26}) with $\mathcal Q_n$ as in (\ref{2.29}). The prolongation of $\hat X_E$ is
\bea \label{2.32}
\mbox{pr} \hat X_E = \sum_j \mathcal Q_j \partial_{u_j} + \sum_{k=1}^K \sum_{j`} (D_t^k \mathcal Q_j)\partial_{u_j^{(k)}},
\eea
where the $j$ summation is over all points figuring in eq. (\ref{2.8}) and $u_j^{(k)}$ denotes the $k$--th $t$--derivative of $u_j$.

The standard vector field $\hat X$ generating Lie point symmetries and its prolongation are given by the formula (\ref{2.30}). Explicitly the prolongation formula is
\bea \label{2.33}
\mbox{pr} \hat X &=& \phi_n \partial_{u_n} + \tau_n \partial_t + \sum_{j \ne n} \phi_j \partial_{u_j}  \\ \nonumber
&+&\sum_j \sum_{k=1}^K \phi_j^{[k]} \partial_{u_j^{[k]}} + \sum_j \sum_{k=1}^K (\tau_n - \tau_j) (D_t^{k+1} u_j) \partial_{u_j^{[k]}},
\\ \label{2.34}
\phi_j^{[k]} &=& D_t \phi_j^{[k-1]} - (D_t \tau_j) u_j^{[k]}, \qquad D_t^k u_j \equiv u_j^{[k]}.
\eea
Notice that $\phi_j^{[k]}$ is the same as for a differential equation \cite{r1} but the last term in (\ref{2.33}) has no analog in the continuous case. The coefficients $\phi_n$ and $\tau_n$ in the vector field $\hat X$ itself are a priori functions of $n$, $t$ and $u_n$ (see eq. (\ref{2.10})). In the following section we will examine some cases when $\tau_n(t,u_n)$ simplifies.

Eq. (\ref{2.33}) is also obtained as the semicontinuous limit of the discrete prolongation (\ref{2.5})

\section{Theorems Simplifying the Calculation of Symmetries of D$\Delta$E.}\label{s3}

\subsection{General comments}

Lie point symmetries of D$\Delta$E of the form (\ref{2.8}) are generated by vector fields of the form (\ref{2.10}). We shall now investigate 3 important cases when the coefficient $\tau_n(t,u_n)$ actually depends on $t$ alone.

The 3 classes of D$\Delta$E are
\bea \label{3.1}
&\dot u_{n}=&f_n(t,u_{n-1}, u_n, u_{n+1}),
\\ \label{3.2}
&\ddot u_{n}=&f_n(t, \dot u_n, u_{n-1}, u_n, u_{n+1}),
\\ \label{3.3}
&u_{n,xy} =&f_n(x,y,u_{n,x}, u_{n,y},u_{n-1}, u_n, u_{n+1}).
\eea
Eq. (\ref{3.1}) contains integrable Volterra, modified Volterra and discrete Burgers type equations \cite{r5}. A list of integrable Toda type equations of the form (\ref{3.2}) can be found in the  reference \cite{y93}. The class (\ref{3.3}) involves 2 continuous variables and contains the two dimensional Toda model \cite{m,fg}.  A list of integrable cases exists \cite{sy97} and Lie point symmetries of this class have been studied.

For equations (\ref{3.1}) and (\ref{3.2}) Lie point symmetries correspond to commuting flows of the form (\ref{2.23}) with $\mathcal Q_n$ given by eq. (\ref{2.29}) while for eq. (\ref{3.3}) the form is
\begin{subequations}\label{3.5}
\bea \label{3.5a}
&&u_{n,\lambda} = \psi_n(x, y, u_n,  u_{n,x}, u_{n,y}),
\\ \label{3.5b}
&&\psi_n(x, y, u_n,  u_{n,x}, u_{n,y}) = \phi_n(x,y,u_n) - \xi_n(x,y,u_n) u_{n,x} - \eta_n(x,y,u_n) u_{n,y}.
\eea
\end{subequations}

For all equations (\ref{3.1}, \ref{3.2}, \ref{3.3}), we assume everywhere below that at least one of the following two conditions is satisfied:
\bea \label{cond}
\frac{\partial f_n}{\partial u_{n+1}} \ne 0 , \quad \hbox{for all} \ n , \qquad \mbox{or} \qquad  \frac{\partial f_n}{\partial u_{n-1}} \ne 0 , \quad \hbox{for all} \ n .
\eea
\subsection{Volterra type equations and their generalizations.}

Let us consider eq. (\ref{3.1}).

\begin{theorem}\label{1} If  (\ref{3.1}) satisfies at least one of the
conditions (\ref{cond}) and (\ref{2.23}) represents a point symmetry of (\ref{3.1}) then we have
\bea \label{3.7}
\tau_n(t,u_n) = \tau(t) .
\eea
\end{theorem}

\begin{proof}
The compatibility condition (\ref{2.24}) of eqs. (\ref{3.1}) and (\ref{2.23}, \ref{2.29}) implies
\bea \label{3.8}
&&\sum_{l=-1}^1 f_{n,u_{n+l}} [\phi_{n+l} - \tau_{n+l} f_{n+l}] + (\tau_{n,t} + \tau_{n,u_{n}} f_n ) f_n +
\\ \nonumber
&& + \tau_n [f_{n,t} + \sum_{l=-1}^1 f_{n,u_{n+l}} f_{n+l}] - \phi_{n,t} - \phi_{n,u_n} f_n = 0 ,
\eea
where indices $t$, $u_{n+l}$ denote partial derivatives.
Taking the derivative of eq. (\ref{3.8}) with respect to $u_{n+2}$ and separately with respect to $u_{n-2}$, we obtain two relations:
\bea \label{3.9}
&&f_{n+1,u_{n+2}} f_{n,u_{n+1}} (\tau_n - \tau_{n+1}) = 0,
\\ \nonumber
&&f_{n-1,u_{n-2}} f_{n,u_{n-1}} (\tau_{n} - \tau_{n-1}) = 0.
\eea
In view of the conditions (\ref{cond}), eqs. (\ref{3.9}) imply
\bea \label{3.10}
\tau_{n+1}(t,u_{n+1}) = \tau_n(t, u_n) \quad \hbox{or} \quad \tau_{n-1}(t,u_{n-1}) = \tau_n(t, u_n) .
\eea
Each of these conditions must be satisfied for any $n$ and they are equivalent.
Since $u_0, u_1, u_{-1}, \dots$ are independent, we find that $\tau_n(t, u_n)$ depends on $t$ alone and this proves Theorem \ref{1}. \qed
\end{proof}

A somewhat weaker theorem can be proved for a more general differential--difference equation, namely,
\bea \label{3.11}
\dot u_n = f_n(t,u_{n+k}, u_{n+k+1}, \cdots, u_{n+m}), \quad k \le  m .
\eea

\begin{theorem} \label{2}
Let Eqs. (\ref{2.23}, \ref{2.29}) represent a symmetry of eq. (\ref{3.11}). If the function $f_n$ in (\ref{3.11}) satisfies
\bea \label{3.12}
m > 0, \qquad \frac{\partial f_n}{\partial u_{n+m}} \ne 0 \quad \hbox{for all} \ n ,
\eea
 then the function $\tau_n(t,u_n)$ is such that
\bea \label{3.13}
\tau_n(t,u_n) = \tau_n(t), \qquad \tau_{n+m}(t) = \tau_n(t).
\eea
If the function $f_n$ satisfies
\bea \label{3.14}
k < 0, \qquad \frac{\partial f_n}{\partial u_{n+k}} \ne 0 \quad \hbox{for all} \ n ,
\eea
then we have
\bea \label{3.15}
\tau_n(t,u_n) = \tau_n(t), \qquad \tau_{n+k}(t) = \tau_n(t).
\eea
\end{theorem}

\begin{proof}
The compatibility condition for Eqs.(\ref{2.23}, \ref{2.29}) and (\ref{3.11}) will be the same as Eq. (\ref{3.8}) but all sums will be from $l=k$ to $l=m$. If Eq. (\ref{3.12}) is satisfied we can differentiate eq. (\ref{3.8}) with respect to $u_{n+2m}$ and obtain $\tau_n(t,u_n)= \tau_{n+m}(t,u_{n+m})$ which implies (\ref{3.13}). If (\ref{3.14}) is satisfied we differentiate (\ref{3.8}) with respect to $u_{n+2k}$ and obtain $\tau_n(t,u_n)= \tau_{n+k}(t,u_{n+k})$ which implies (\ref{3.15}). \qed
\end{proof}

This result is valid, in particular, for Burgers type equations for which $k=0$, $m>0$ or $k<0$, $m=0$
For all equations in the class (\ref{3.11}), under the assumptions of this theorem,  the function $\tau_n$ is independent of $u_n$ and is periodic in $n$. In particular, if $k=-2$, $m=2$  it is two-periodic and we can write
\bea \label{3.16}
\tau_n(t) = \frac{1+(-1)^n}{2} \tau_0(t) + \frac{1-(-1)^n}{2} \tau_1(t).
\eea

\subsection{Toda type equations}

The compatibility condition for eq. (\ref{3.2}) and Eqs. (\ref{2.23}, \ref{2.29}) is $ u_{n,tt\lambda} =  u_{n,\lambda tt}$ and implies
\bea \label{3.17}
&&f_{n,\dot u_n} [ \phi_{n,t}+(\phi_{n,u_n} - \tau_{n,t})\dot u_{n}-\tau_{n,u_n} (\dot u_{n})^2 ]
\\ \nonumber
&& + \sum_{k=-1}^1 f_{n,u_{n+k}} [ \phi_{n+k} + (\tau_n - \tau_{n+k}) \dot u_{n+k} ]
\\ \nonumber
&&- \phi_{n,tt} + (-2 \phi_{n,t u_n} + \tau_{n,tt}) \dot u_{n} + (-\phi_{n, u_n u_n} + 2 \tau_{n,t u_n }) (\dot u_{n})^2
\\ \nonumber
&&+ \tau_{n, u_n u_n} (\dot u_{n})^3 + \tau_n f_{n,t} + (2 \tau_{n,t} - \phi_{n, u_n} + 3 \tau_{n,u_n} \dot u_{n})f_n = 0.
\eea
We use Eq. (\ref{3.17}) to prove the following theorem.

\begin{theorem} \label{3}
Let Eqs. (\ref{2.23}, \ref{2.29}) represent a point symmetry of Eq. (\ref{3.2}) and let the function $f_n$ in Eq. (\ref{3.2}) satisfy at least one of the conditions (\ref{cond}) for all $n$. Then the function $\tau_n(t,u_n)$ in eq. (\ref{2.29}) satisfies Eq. (\ref{3.7}), i.e. $\tau_n(t, u_n)$ depends on $t$ alone.
\end{theorem}

\begin{proof}
None of the functions $f_n$, $\phi_n$, $\tau_n$ figuring in eq. (\ref{3.17}) depends on $\dot u_{n+1}$ or $\dot u_{n-1}$. These two expressions do however figure explicitly in (\ref{3.17}). Their coefficients must hence vanish and we obtain
\bea \label{3.18}
&& f_{n,u_{n+1}}(\tau_n - \tau_{n+1}) = 0,
\\ \nonumber
&& f_{n,u_{n-1}}(\tau_n - \tau_{n-1}) = 0.
\eea
In view of the conditions (\ref{cond}), we can use one of eqs. (\ref{3.18}), and both of them provide the same:
\bea \label{3.18a}
\tau_n(t,u_n) = \tau_{n+1}(t,u_{n+1})
\eea
for any $n$. Hence we again obtain the result (\ref{3.7}), as stated in Theorem \ref{3}.  \qed
\end{proof}

\subsection{Toda field theory type equations}

Let us now consider the equation (\ref{3.3}) and assume that it has a Lie point symmetry represented by (\ref{3.5}).
\begin{theorem} \label{4}
Let (\ref{3.5}) represent a Lie point symmetry of the field equation (\ref{3.3}) and let the function $f_n(x,y,u_{n,x}, u_{n,y}, u_{n-1}, u_n, u_{n+1})$ satisfy at least one of the conditions
\bea \label{3.19}
\frac{\partial f_n}{\partial u_{n-1}} \ne 0, \quad \mbox{or} \quad  \frac{\partial f_n}{\partial u_{n+1}} \ne 0.
\eea
The functions $\xi_n$ and $\eta_n$ in the symmetry (\ref{3.5}) then are given by
\bea \label{3.20}
\xi_n(x,y,u_n)=\xi(x,y), \qquad \eta_n(x,y,u_n)=\eta(x,y).
\eea
\end{theorem}
 \begin{proof}
The compatibility condition $u_{n, x y \lambda}=u_{n,\lambda x y}$ in this case can be written as
\bea \label{3.21}
\sum_{k=-1}^1 \frac{\partial f_n}{\partial u_{n+k}} \psi_{n+k} + \frac{\partial f_n}{\partial u_{n,x}} D_x \psi_n + \frac{\partial f_n}{\partial u_{n,y}} D_y \psi_n - D_x D_y \psi_n = 0
\eea
with $\psi_n$ as in Eq. (\ref{3.5b}); $D_x$ and $D_y$ are the total derivative operators. The terms $u_{n \pm 1,x}$, $u_{n \pm 1,y}$ only figure in $\psi_{n \pm 1}$ and in $D_x D_y \psi_n$ where we have
\bea \nonumber
D_x D_y \psi_n = - \xi_n ( D_x f_n) - \eta_n (D_y f_n) + \cdots
\eea
with
\bea \nonumber
D_x f_n = \frac{\partial f_n}{\partial u_{n-1}} u_{n-1,x} + \frac{\partial f_n}{\partial u_{n+1}}u_{n+1,x} + \cdots
\\ \nonumber
D_y f_n = \frac{\partial f_n}{\partial u_{n-1}} u_{n-1,y} + \frac{\partial f_n}{\partial u_{n+1}} u_{n+1,y} + \cdots .
\eea
Substituting into (\ref{3.21}) and setting the coefficients of $u_{n \pm 1, x}$ and $u_{n \pm 1, y}$ equal to zero separately, we obtain
\bea \label{3.22}
&(\xi_{n-1}-\xi_n) \frac{\partial f_n}{\partial u_{n-1}} = 0, \quad &(\eta_{n-1}-\eta_n) \frac{\partial f_n}{\partial u_{n-1} }= 0,
\\ \nonumber
&(\xi_{n+1}-\xi_n) \frac{\partial f_n}{\partial u_{n+1}} = 0, \quad &(\eta_{n+1}-\eta_n) \frac{\partial f_n}{\partial u_{n+1}} = 0.
\eea
Thus, under the assumption (\ref{3.19}) we obtain (\ref{3.20}) and this completes the proof.$\quad$\qed
\end{proof}

\section{Examples}\label{s4}
Let us now consider examples of each of the classes of differential--difference equations discussed in Section 3.

\subsection{The YdKN equation}

The Krichever--Novikov equation \cite{1}
\bea \label{71a}
\dot u = \frac{1}{4} u_{xxx} - \frac{3}{8} \frac{u_{xx}^2}{u_x} + \frac{3}{2} \frac{P(u)}{u_x},
\eea
where $P(u)$ is an arbitrary fourth degree polynomial with constant coefficients, is an integrable PDE with many interesting properties \cite{1,3,5,lpsy,kn,7,9,6,8,4,asy}.

Yamilov and collaborators have proposed integrable discretizations of eq. (\ref{71a}) \cite{r51,r5,y06}. The original form of the YdKN equation \cite{r5,y06} is
\bea \label{72a}
u_{n,t} &=& \frac{P_n u_{n+1}u_{n-1} + Q_n ( u_{n+1} + u_{n-1} ) + R_n}{u_{n+1} - u_{n-1}}, \\ \label{73a}
P_n &=& \alpha u_n^2 + 2 \beta u_n + \gamma, \\ \nonumber
Q_n &=& \beta u_n^2 + \lambda u_n + \delta, \\ \nonumber
R_n &=& \gamma u_n^2 + 2 \delta u_n + \omega,
\eea
where $\alpha, \cdots , \omega$ are pure constants.

A complete symmetry analysis of this equation and its generalizations is in preparation \cite{LWY2010}. Here we will just consider one special case as an example of a Volterra type equation.  Let us set $\alpha=1$, $\beta= \cdots = \omega =0$ in (\ref{73a}). The YdKN equation reduces to
\bea \label{74a}
u_{n,t} = \frac{u_n^2 u_{n+1} u_{n-1}}{u_{n+1} - u_{n-1}}.
\eea
According to Theorem 3.1 a compatible flow corresponding to a point symmetry will have the form
\bea \label{75a}
u_{n,\lambda} = \Phi_n(t, u_n) - \tau(t) u_{n,t}.
\eea
We replace $u_{n,t}$ in (\ref{75a}) using (\ref{74a}) and then impose the compatibility condition $u_{n,t\lambda}=u_{n,\lambda t}$. First of all, from terms containing $u_{n+2}$ and $u_{n-2}$ we find that $\Phi_n$ and $\tau$ must satisfy
\bea \label{76a}
\tau &= \tau_0 + \tau_1 t, \quad \Phi_n &= a_n + b_n u_n + c_n u_n^2, \\ \nonumber
a_n &= a + \hat a (-1)^n, \quad b_n &= b + \hat b (-1)^n, \quad c_n = c + \hat c (-1)^n,
\eea
where $\tau_0, \tau_1, a, \hat a, b, \hat b,$ and $c, \hat c$ are pure constants. This is actually the case for the general YdKN equation (\ref{72a}). Substituting (\ref{76a}) into the compatibility condition we obtain an equation that is polynomial in $u_{n+k}$. Setting  coefficients of $u_{n-1}^a u_{n}^b u_{n+1}^c$ equal to zero for each independent term we obtain the following basis of the Lie point symmetry algebra of eq. (\ref{74a})
\bea \label{77a}
X_1 &=& \partial_t, \quad X_2 = u_n^2 \partial_{u_n}, \quad X_3 = (-1)^n u_n^2 \partial_{u_n}, \\ \nonumber
X_4 &=& t \partial_t - \frac{1}{2} u_n \partial_{u_n}, \qquad X_5 = (-1)^n u_n \partial_{u_n}.
\eea

This is a solvable Lie algebra with $\{ X_1, X_2, X_3 \}$its Abelian niilradical. The two nonnilpotent elements satisfy $[X_4, X_5]=0$ and their action on the nilradical is given by
\bea \label{78a}
 \left(
\begin{array}{c}
{[X_4,X_1]}\\
{[X_4,X_2]}\\
{[X_4,X_3]}
\end{array}
\right)&=&\left(
\begin{array}{ccc}
-1&0&0\\
0&-\frac{1}{2}&0 \\
0&0& -\frac{1}{2}
\end{array}
\right)  \left(
\begin{array}{c}
X_1\\
X_2 \\
X_3
\end{array}
\right) , \\ \nonumber  \left(
\begin{array}{c}
{[X_5,X_1]}\\
{[X_5,X_2]} \\
{[X_5,X_3]}
\end{array}\right)&=&\left(
\begin{array}{ccc}
0&0&0\\
0&0&1 \\
0&1&0
\end{array}
\right)  \left(
\begin{array}{c}
X_1\\
X_2 \\
X_3
\end{array}
\right).
\eea

\subsection{The Toda lattice}
The Toda lattice itself
\bea \label{79a}
u_{n,tt} = \exp(u_{n-1}-u_n) - \exp(u_n-u_{n+1}),
\eea
is the best known example of an equation of the type (\ref{3.2}). According to Theorem 3.3 the flow corresponding to its point symmetries will satisfy (\ref{75a}). From the compatibility condition $u_{n,tt\lambda}=u_{n,\lambda tt}$ we obtain the Lie point symmetry algebra
\bea \label{80a}
X_1 = \partial_t, \quad X_2 = t \partial_{u_n}, \quad X_3 = \partial_{u_n}, \quad X_4 = t \partial_t + 2 n \partial_{u_n}.
\eea
This Lie algebra is solvable, its nilradical $\{ X_1, X_2, X_3 \}$ is isomorphic to the Heisenberg algebra. We note that the Ansatz made in \cite{r4} was not correct and lead to $X_3 = q(n) \partial_{u_n}$ in (\ref{80a}) with $q(n)$ arbitrary. It was however noted there that a closed Lie algebra is obtained only for $q(n) = \mbox{const}$.

\subsection{The two--dimensional Toda lattice equation.}

The equation to be considered \cite{fg,m} is
\bea \label{81a}
u_{n,xy} =  \exp(u_{n-1}-u_n) - \exp(u_n-u_{n+1}).
\eea
According to Teorem 3.4 the flow corresponding to point symmetries will take the form
\bea \label{82a}
u_{n,\lambda} = \phi_n(x,y,u_n) - \xi(x,y) u_{n,x} - \eta(x,y) u_{n,y}.
\eea
The Lie point symmetry algebra obtained from the compatibility condition $u_{n,xy\lambda} = u_{n,\lambda xy}$ is infinite--dimensional and depends on 4 arbitrary function of one variable each
\bea \label{83a}
X(f) &=& f(x) \partial_x + f'(x)n \partial_{u_n}, \quad  U(k) = k(x) \partial_{u_n}, \\ \nonumber
X(g) &=& g(y) \partial_y + g'(y)n \partial_{u_n}, \quad W(\ell) = \ell(y) \partial_{u_n}.
\eea
This algebra happens to coincide with the one found in  \cite{r41} through the prolongation formula used there was incorrect. This is a Kac--Moody--Virasoro algebra as is typical for integrable equations with more than 2 independent variables (in this case $x,y$ and $n$).

\section{Conclusions}

The main results of the present article are:
\begin{enumerate}
\item The prolongation formulas (\ref{2.31}) and (\ref{2.32}), (\ref{2.33}) for evolutionary and ordinary vector fields generating commuting flows and Lie point symmetry transformation for differential--difference equations. These are viewed here as differential equations on fixed non--transforming lattices.
\item The prolongation formulas and the corresponding algorithm for calculating Lie point symmetries of differential--difference equations are greatly simplified for 3 rather general classes of equations ( including the Toda lattice, the two--dimensional Toda lattice and the Volterra equations). The results are summed up in Theorems 3.1--3.4.
\item We have presented an example of each class of equations covered by the above Theorems and identified a class of equations depending on 6 parameters (with generalizations depending on 9  parameters). These are the Yamilov discretizations of the Krichever--Novikov equation (\ref{72a}, \ref{73a}).

\end{enumerate}

A complete analysis of the symmetries of the YdKN equation and its generalizations will be published separately.

\paragraph{Acknowledgments.}
RIY has been partially supported by the Russian Foundation for Basic Research (grant
numbers 08-01-00440-a and 09-01-92431-KE-a).   The research of P.W. was partially supported by a research grant from
NSERC of Canada. LD has been partly supported by the Italian Ministry of Education and Research, PRIN
"Nonlinear waves: integrable fine dimensional reductions and discretizations" from 2007
to 2009 and PRIN "Continuous and discrete nonlinear integrable evolutions: from water
waves to symplectic maps" from 2010.


\begin{thebibliography}{99} \small



\bibitem{asy} V.E. Adler, A.B. Shabat and R.I. Yamilov, Symmetry approach
to the integrability problem, {\it Teoret. Mat. Fiz.} {\bf 125} (2000), no. 3, 355--424 (in
Russian); English translation in {\it Theor. Math. Phys.} {\bf 125} (2000), no. 3, 1603--1661.

\bibitem{r2} V.A. Dorodnitsyn {\it The Group Properties of Difference Equations} (Moscow, Fizmatlit, 2001) (in Russian).

\bibitem{5}B. Dubrovin, I. Krichever, and S. Novikov, "Integrable systems. I," in: Itogi Nauki i Tekhn. Ser.: Sovr. Probl.
Matem. [in Russian] (R. V. Gamkrelidze, ed.), Vol. 4, Dynamic Systems-4, VINITI, Moscow (1985), pp. 179-285.

\bibitem{fg} A.P. Fordy and J. Gibbons {\it Integrable nonlinear Klein--Gordon equations and Toda lattices} Comm. Math. Phys. {\bf 71} (1980) 21--30.

\bibitem{1} 1. I. M. Krichever and S. P. Novikov,Holomorphic bundles and nonlinear equations. Finite-gap solutions of rank $2$, {\it  Sov. Math. Dokl.}  {\bf 20} (1979) 650-654.

\bibitem{kn} I.M. Krichever and S.P. Novikov, Holomorphic bundles over algebraic curves and non-linear equations, {\it Uspekhi Mat. Nauk} {\bf 35} (1980), 47--68 (in Russian); English translation in {\it Russ. Math. Surv.} {\bf 35} (1980), 53--80.

\bibitem{8}G. Latham and E. Previato,Darboux transformations for higher-rank Kadomtsev-Petviashvili and Krichever-Novikov equations. {\it Acta Appl. Math.}, {\bf  39} (1995) 405-433.

\bibitem{lpsy} D. Levi, M. Petrera, C. Scimiterna and R. Yamilov, On Miura transformations and Volterra-type equations associated with the Adler-Bobenko-Suris equations, {\it SIGMA} {\bf 4} (2008), 077, 14 pages.

\bibitem{r4}D. Levi and P. Winternitz, Continuous symmetries of discrete equations.  {\it Phys. Lett.} {\bf A  152}  (1991) 335--338.

\bibitem{r41}D. Levi and P.  Winternitz, Symmetries and conditional symmetries of differential-difference equations.  {\it J. Math. Phys. } {\bf  34}  (1993) 3713--3730.

\bibitem{r3}D. Levi and P. Winternitz, Continuous symmetries of difference equations, {\it J. Phys. A: Math. Gen.} {\bf 39} (2006) R1–-R63.

\bibitem{LWY2010} D. Levi, P. Winternitz and R. Yamilov, {\it Point Symmetries of the Yamilov Discretization of the
Krichever-Novikov
Equation and its Generalizations} in preparation.

\bibitem{r51} D. Levi and R. Yamilov, {\it Conditions for the existence of
higher symmetries of evolutionary equations on the lattice}, J. Math. Phys.
{\bf 38} (1997),  6648--6674.

\bibitem{m} A.V. Mikhailov, {\it Integrability of two dimensional Toda chain} Sov. Phys. JETP Lett. {\bf 30} (1979) 414--418.

\bibitem{7}O. I. Mokhov, Canonical Hamiltonian representation of the Krichever-Novikov equation, {\it Math. Notes}, 50, 939-945 (1991).

\bibitem{9}D. P. Novikov, {\it Algebraic--geometric solutions of the Krichever--Novikov equation}, Theoretical and Mathematical Physics, {\bf  121},  1999 1567--1573.

\bibitem{3}S. P. Novikov, S. V. Manakov, L. P. Pitaevsky, and V. E. Zakharov, Theory of Solitons: The Inverse Scattering
Method [in Russian], Nauka, Moscow (1980); English transl., Plenum, New York (1984).

\bibitem{r1} P.J. Olver {\it Applications of Lie Groups to Difference Equations} (New York, Springer 2000).

\bibitem{sy97} A.B. Shabat and R.I. Yamilov, {\it To a transformation theory of
two-dimensional integrable systems},  Phys. Lett. A {\bf 227} (1997) 15--23.

\bibitem{6}V. V. Sokolov, Hamiltonian property of the Krichever-- Novikov equation {\it Sov. Math. Dokl.}, {\bf 30}  (1984) 44-46.

\bibitem{4}S. I. Svinolupov, V. V. Sokolov, and R. I. Yamilov,  B\"acklund transformations for integrable evolution equations. {\it Sov. Math. Dokl.}, {\bf 28} (1983) 165-168.

\bibitem{r31} P. Winternitz, Symmetries of Discrete Systems, {\it Discrete Integrable Systems} edited by B. Grammaticos, Y. Kosmann--Schwarzbach and T. Tamizhmani (Berlin, Springer, 2004) p. 185--243.

\bibitem{r5} R.I. Yamilov, Classification of discrete evolution equations,
{\it Uspekhi Mat. Nauk} {\bf 38} (1983), no. 6, 155--156 (in  Russian).

\bibitem{y93} R.I. Yamilov, Classification of Toda type scalar
lattices, In: Proceedings of Int. Workshop NEEDS'92
(eds: V. Makhankov, I. Puzynin, O. Pashaev), World Scientific Publishing,
1993, 423--431.

\bibitem{y06} R. Yamilov, Symmetries as integrability criteria for
differential difference equations, {\it J. Phys. A: Math. Gen.} {\bf 39} (2006)
R541--R623.

\end{thebibliography}
 \end{document}